\documentclass[conference]{IEEEtran}
\usepackage{color}
\usepackage[english]{babel}
\usepackage{graphicx}
\usepackage{epstopdf}
\usepackage{times}
\usepackage{multirow}
\usepackage[algo2e,linesnumbered]{algorithm2e} 
\usepackage{algorithmic}
\usepackage{algorithm}
\usepackage{psfrag}
\usepackage[shell]{gnuplottex}
\usepackage{mathenv}
\usepackage{array}
\usepackage[T1]{fontenc}
\usepackage{array}
\usepackage{mdwmath}
\usepackage{multicol}
\usepackage{dblfloatfix}
\usepackage{amsmath}
\usepackage{amssymb}
\usepackage{float}
\usepackage{url}

\makeatletter
\newcommand*{\rom}[1]{\expandafter\@slowromancap\romannumeral #1@}
\makeatother

\begin{document}

\long\def\/*#1*/{}
\title{ Dynamic Channel Access Scheme for Interference Mitigation in Relay-assisted Intra-WBANs}

  \author{\IEEEauthorblockN{Mohamad Jaafar Ali\IEEEauthorrefmark{1}, Hassine Moungla\IEEEauthorrefmark{1}, Ahmed Mehaoua\IEEEauthorrefmark{1}}
\IEEEauthorblockA{\IEEEauthorrefmark{1}LIPADE, University of Paris Descartes, Sorbonne Paris Cit\'{e},
45 rue des saints p\`{e}res, 75006, Paris, France \\
Email: \{mohamad.ali; hassine.moungla; ahmed.mehaoua\}@parisdescartes.fr}}

\maketitle
\begin{abstract}
This work addresses problems related to interference mitigation in a single wireless body area network (WBAN). In this paper, We propose a distributed \textit{C}ombined carrier sense multiple access with collision avoidance (CSMA/CA) with \textit{F}lexible time division multiple access (\textit{T}DMA) scheme for \textit{I}nterference \textit{M}itigation in relay-assisted intra-WBAN, namely, CFTIM. In CFTIM scheme, non interfering sources (transmitters) use CSMA/CA to communicate with relays. Whilst, high interfering sources and best relays use flexible TDMA to communicate with coordinator (C) through using stable channels. Simulation results of the proposed scheme are compared to other schemes and consequently CFTIM scheme outperforms in all cases. These results prove that the proposed scheme mitigates interference, extends WBAN energy lifetime and improves the throughput. To further reduce the interference level, we analytically show that the outage probability can be effectively reduced to the minimal.
\end{abstract}
\IEEEpeerreviewmaketitle

\section{Introduction}
A WBAN is a wireless network of low power and low cost sensors that may be embedded inside or attached on the human body. Its sensors are used in various applications such as personal health monitoring, ubiquitous healthcare, sports and military. It mainly monitors and captures vital signs as glucose percentage in blood, heart beats, respiration and/or can record electrocardiography (ECG) \cite{key2,key14,key15,key19}.

Due to the dynamic and mobile nature of WBANs, in TDMA access scheme, interferences can happen due to the infeasibility of coordination among different coexisting WBANs (inter-WBANs interference). Or, in CSMA/CA access scheme, it can happen when multiple nodes of a particular WBAN can not concurrently transmit (intra-WBAN interference). At the first side, radio co-channel interference is of the paramount importance. It increases energy consumption. Further, it also decreases the efficiency of the reliable communication and the throughput. On the other side, the stringent factor is energy and this requires to keep always as low power consumption as possible.

Recently, the IEEE 802.15.6 working group has defined new proposals for WBANs and also adopted the cooperative two-hop scheme in the standard \cite{key26}. Thus, adopting relay transmission scheme is a very promising solution for co-channel interference mitigation \cite{key21,key8,key10,key17,key7,key22}. Firstly, co-channel interference problem motivates for the stringent requirements of interference mitigation and/or avoidance schemes, protocols for reliable and energy efficient operation of both isolated and coexisting WBANs. Secondly, due to the constrained nature of WBANs (in terms of energy, size and cost), advanced antenna techniques cannot be used for interference mitigation as well as power control mechanisms used in cellular networks are not applicable to WBANs \cite{key21,key22,key15}.

However, in this work we stress our concentration on problems related to co-channel interference and power savings of an isolated WBAN. Thus, novel methods and schemes are required for intra-WBAN interference mitigation and consequently for energy savings. 

The rest of the paper is organized as follows. Section \rom{2} shows the works address problems related to interference mitigation in WBANs. Section \rom{3} explains the flexible TDMA scheme. Section \rom{4} presents the system model and introduces the proposed CFTIM scheme. Section \rom{5} shows the simulation and comparison results. The conclusions are drawn in section \rom{6}. 

\section{Related Work}
Recent studies show multihop schemes have a lower power consumption in comparison to one-hop scheme. However, using relays reduces the WBAN interference and consequently the power consumption. Authors propose a single-relay cooperative scheme where the best relay is selected in a distributed fashion \cite{key15, key14}. Also, authors of \cite{key17} propose a prediction-based dynamic relay transmission scheme through which the problem of "when to relay" and "who to relay" are decided in an optimal way. The interference problem among multiple co-located WBANs is investigated in \cite{key7}. The authors show cooperative two relay communication with opportunistic relaying significantly mitigates WBAN interference.

Authors of \cite{key8} propose an analytical framework to optimize the size of relay-zone around each source node. Authors of \cite{key22} investigate the problem of coexistence of multiple non coordinated WBANs. This study provides better co-channel interference mitigation. However, more recent works conducted in \cite{key9} propose a scheme for joint two-hop relay-assisted cooperative communication integrated with transmit power control. This scheme can reduce co-channel interference and extend the lifetime. 

On the other hand, most of the recent works prove that TDMA scheme is an attractive solution to avoid interference inside a small-sized WBAN \cite{key2, key14, key15, key19, key20, key21}. Furthermore, authors of \cite{key6} also enables two or three coexisting WBANs to agree on a common TDMA schedule to reduce the interference. The work in \cite{key3} adopts a TDMA polling-based scheme for traffic coordination within a WBAN and a carrier sensing (CS) mechanism to deal with inter-WBAN interference.

Some efforts develop a model to analyze interference by using geometrical probability approach among non-overlapping nearby WBANs \cite{key4}. Other research focuses on the performance at the coordinator that calculates signal-to-interference-plus-noise ratio (SINR) periodically which enables to command its nodes to select appropriate interferece mitigation scheme \cite{key1}. Other studies of \cite{key6} analyze the performance of a reference WBAN. They evaluate the performance in terms of bit error rate, throughput and lifetime which have been improved by adoption of an optimized time hopping code assignment strategy. Works in \cite{key5} consider a WBAN where coordinator periodically queries sensors to transmit data. The network adopts the CSMA/CA and the nodes adopt link adaptation to select the modulation scheme according to the experienced channel quality \cite{key14,key15, key19}.

Whereas, the research work of \cite{key23} solves the problem of inter-WBAN scheduling and interference by the adoption of a QoS based MAC preemptive priority scheduling approach. Whilst, researchers of \cite{key20} proposes a distributed interference detection and mitigation scheme through using adaptive channel hopping. Research work of \cite{key16} proposes a dynamic resource allocation scheme for interference avoidance among multiple coexisting WBANs through using orthogonal sub-channels for high interfering nodes.

Most of the recent works show problems related to interference mitigation in inter-WBANs environment. They do not address problems related to interference and energy savings of a single isolated WBAN. In this paper, we propose a distributed relay-assisted algorithm for interference mitigation within a single WBAN. The proposed scheme enables non interfering sources (transmitters) to use CSMA/CA for transmitting to relays. Whilst, high interfering sources and best relays transmit to the coordinator directly through using stable channels in their assigned time slots.

\section{Proposed Flexible TDMA Scheme (FTDMA)} 
In the traditional TDMA scheme, a superframe is usually splitted into a number of equal intervals each called timeslot. Each timeslot is assigned to a node to transmit its message to C. A proposed TDMA scheduling ensures collision-free communication. However, in WBANs, nodes sleep and wake up dynamically and hence, the number of transmitting nodes is unexpected. Therefore, a dynamic and flexible way of scheduling different transmissions is required to efficiently avoid interference, save energy and decrease the time delay. 
\subsection{The FTDMA Frame Structure}
\begin{figure}[ht]
  \centering
        \includegraphics[width=0.5\textwidth]{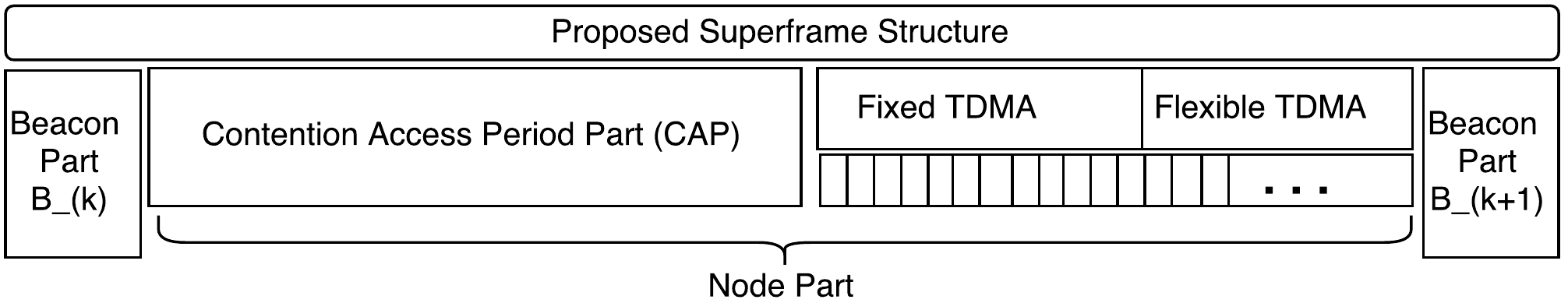}
\caption{Proposed CFIP Superframe Structure}
  \label{frame}
\end{figure}

Unlike the traditional TDMA scheme, a FTDMA frame is divided into two parts. The first part is used by C for broadcasting beacons to the nodes and the second part is used by the nodes for transmitting their messages to C. However, this latter part is further splitted into a CSMA/CA part and a TDMA part which its size changes dynamically depending on the interference level. See Fig. \ref{frame}.  

In the C part of a FTDMA frame, C boadcasts the information based on the messages received from its nodes during the node part of the previous frame. Along with this information, it allocates a slot in the subsequent node part for each active node. However, a node is considered active if C has received at least one beacon during the previous three frames. If there are m active nodes currently connected to C, C allocates p (where p > m) slots in the subsequent node part. The first m slots are allocated to the currently active nodes and the rest p-m slots are reserved to the newly incoming nodes (those nodes have sensed data and do not have assigned timeslots in the current superframe). However, C predicts p for the next frame based on the interference level history of the last k-received frames where k can be any interger.

The node part is composed of two main parts, a CSMA/CA part denoted by CAP and a TDMA part as shown in Fig. \ref{frame}. In the CAP, non interfering sources transmit their message to relays. Whilst, in the TDMA part, a node (can be a high interfering source or a best relay sensor) first listens to the beacon from C. If it finds its ID in the slot allocation list of the beacon, it transmits its message in the corresponding timeslot (one of the m slots). However, if it does not find its ID in the slot allocation list. It synchronizes its clock with that of C. It then randomly selects one of the p-m empty slots and transmits its beacon meassage in that slot. Since these slots are free and not yet assigned, there is chance of collision with the transmission of any other node willing to trnasmit at the same slot. If the beacon is successfully sent to C, i.e. there is no collision, C will allocate a slot for the node in the next frame. Otherwise, if collision happens, C will not successfully receive the beacon, and will not include its ID in the next frame. In such cases, a node keeps trying different empty slots randomly in every frame until a timeslot is assigned to it. 

\subsection{A Mechanism to Handle with Different Interference Levels}
In our proposed FTDMA, C manages the incoming messages from the nodes according to the interference level by varying the number of empty slots only (p-m). Therefore, when the interference is low, few empty slots can handle the newly incoming nodes. If C finds that many empty slots remain unused, it then lowers the value of p. Furthermore, it is also guanteed that there will be no collision among the messages sent by the nodes since C carefully assigns one slot for one node only. But, there may be collisions in the empty slots reserved for the newly incoming nodes. When the interference increases, more than one node may select the same timeslot causing a collision. Hence, whenever collisions occur, C increases the number of empty slots simply by increasing the value of p.
\section{System Model}

\subsection{Model Assumptions}
The main objective behind our proposal is to $(1)$ minimize interfereces, $(2)$ extend WBAN energy lifetime and $(3)$ maximize the throughput in an intra-WBAN. We define the following:
\begin{itemize}
 \item Non interfering source: a CSMA/CA source has sensed data and whose measured \textbf{SINR} $\geq$ $\textbf{SINR}_{\textbf{Threshold}}$, i.e., channel is clear to transmit.
 \item High interfering source: a CSMA/CA source has sensed data and whose whose measured \textbf{SINR} $<$ $\textbf{SINR}_{\textbf{Threshold}}$, i.e., channel is unclear to transmit.
\end{itemize}
We adopt the following assumptions:
\begin{itemize}
  \item Fixed topology of WBAN nodes (number of interfering nodes is variable from a frame to another depending on interference level)
  \item Topology with combined one- and two-hop as a communication scheme respectively (high interfering sources -> C) and (non interfering sources -> relays -> C)
  \item A combination of CSMA/CA (non interfering sources -> relays)  and flexible TDMA ((high interfering sources -> C) and (best relays -> C)) are adopted as medium access schemes
  \item Any node transmits one data packet in each superframe
  \item The number of available stable channels is always larger than the number of nodes demanding for that channels
  \item One and only one best relay retransmits the source's message to C, i.e. a source can have only one best relay
 \end{itemize}

\subsection{Proposed Scheme}
After the last beacon frame is received from C. All the sources of the WBAN start a CSMA/CA contention to access a shared base channel. During this contention, two sets of sources are composed. non interfering sources denoted by set \textbf{TS} and high interfering sources denoted by set \textbf{IS}. Each source whose \textbf{SINR} $\geq$ $\textbf{SINR}_{\textbf{Threshold}}$ is included in TS. That implies, any member of TS gains the contention to access the channel and transmit its message to the best relay during the current frame. Then, when the flexible TDMA schedule commences, the best relay transmits the message it has received from a member of TS to C. More clearly, a best relay checks the last beacon already received. If it finds its ID in the node slot list (nodeSlotList of m slots) of the beacon. It transmits its message in the corresponding timeslot to C. 
\IncMargin{1em}
\begin{algorithm}
\footnotesize
\SetKwData{Left}{left}\SetKwData{This}{this}\SetKwData{Up}{up}
\SetKwFunction{Union}{Union}\SetKwFunction{FindCompress}{FindCompress}
\SetKwInOut{Input}{input}\SetKwInOut{Output}{output}
\Input{All sources (N), all relays (R), non interfering sources (TS), high interfering sources (IS), all best relays (BR), any source ($s_{i}$), a relay $r_{i}$, a high interfering source or best relay ($rs_{j}$) }

BeginCAP

       \For{$i$ $\leftarrow$ $1$  $\KwTo$  $sizeof(TS)$}
         {

                               $s_{i}$ $\in$ TS transmits to $r_{i}$ $\in$ BR on baseChannel
                            
         }
            
            EndCAP
                
                BeginFTDMA
                            
        \For{$j$ $\leftarrow$ $1$  $\KwTo$  $sizeof(ISBR)$}
          {

               \If {$ID_{rs_{j}}$ $\in$ nodeSlotList of currentFrame} 
                 {
                  
                   $rs_{j}$ $\in$ ISBR transmits to C on stableChannel during $T_{j}$ 
                   
                   C includes $ID_{rs_{j}}$ in nodeSlotList of nextFrame
               
                 }
              
              \Else
              {
                    
                                 counter = 0;
                            
                             \For{$k$ $\leftarrow$ $1$  $\KwTo$  $maxRetries$}
                                 {
          
                                 counter++;
          
                                 $rs_{j}$ $\in$ ISBR selects timeslot $T_{k}$ at random from freeSlotList of currentFrame
                             
                                 \If {$T_{k}$ == freeSlot} 
                                  {
                              
                                    $rs_{j}$ $\in$ ISBR transmits to C on stableChannel during $T_{k}$ 
                                    
                                    C includes $ID_{rs_{j}}$ in nodeSlotList of nextFrame
                                    
                                    counter = 0;
                                    
                                    break;
                              
                                  }
                                  
                                   \If {counter == maxRetries} 
                                    {

                                       $rs_{j}$ $\in$ ISBR  waits for the nextFrame
                                       
                                       counter = 0;
                                   
                                       break;
                                  
                                    }
          
                            }
                
                }
                
          }
          
     EndFTDMA
          
\caption{CFTIM for Intra-WBAN Interference Mitigation}
\label{multihop}
\end{algorithm}
\DecMargin{1em}

However, if it does not find its ID, it then randomly selects one slot of the p-m free slots of free slot list (freeSlotList of p-m slots) and transmits its meassage to C in that slot. 

We denote the set of all IS sources and all best relays by ISBR, $ISBR = IS \cup BR$. The following pseudocode shown in Algo. \ref{multihop} presents the proposed CFTIM scheme and introduces the actions taken at the TS nodes, IS nodes and the best relays. 

On the other hand, high interfering sources whose \textbf{SINR} < $\textbf{SINR}_{\textbf{Threshold}}$ are included to set \textbf{IS}. These sources can not access shared channel during the current frame. This is due to the interference coming from other sources accessing the channel at the same time. Similarly, any member of IS checks the last beacon received from C. If it finds its ID in the node slot list, it transmits its message in the corresponding timeslot directly to C through the most stable channel. However, if it does not find its ID. It then randomly selects one slot of free slot list and transmits its message to C in that slot. If it does not succeed to transmit, it keeps trying different empty slots randomly until a timeslot is assigned to it.

As a node gets aware of its time slot in the current frame, it then starts immediatly scanning a fixed number of channels. Based on the aforementioned assumptions, it finds a stable channel, it switches to this channel and then transmits its message directly to C during its assigned time slot. 

After the flxible TDMA schedule is over, C receives all the IDs of nodes have successfully transmitted in the current frame (say m IDs). However, based on that knowledge, it forms a new beacon frame composed of p slots. C assigns a time slot in the next beacon frame for each node it owns its ID. Consequently, it assigns m slots for m nodes and thus forming the nodeSlotList (fixed TDMA) part of the beacon. Furthermore, it adds p-m unassigned free slots to the beacon and thus forming the freeSlotList (flexible TDMA) part of the beacon. This addition of p-m slots is to let other unassigned nodes to transmit in the next frame. The following pseudocode shown in Algo. \ref{cactions} introduces C actions and transmission phase. The number of time slots specified in the frame is variable from one beacon frame to another thus depending on the interference level.
\subsection{Stability Condition}
We say that a channel is stable with a given probability $P_{thr}$, if its quality is not expected to deteriorate before completing the scheduled data transmission on that channel. More specifically, assume that we are currently in the slot n and we want to select a channel to be used after K time slots. Then, the stability condition for the channel $d_{K+n}$ is:

\begin{equation}
P_{\mu}(i, i) + P_{\mu}(i, i - 1) > P_{thr}
\end{equation}
With
\begin{equation}
\mu = K \times T + T_{data} - x \times \tau_{s}
\end{equation}
Where $T_{data}$ is the duration of the data transmission, x $\times$ $\tau_{s}$ is the total time spent in sensing before finding channel $d_{K+n}$, and i is the cuurent state of channel $d_{K+n}$. in the above condition, the values of interest for i are 1 and 2, because if the channel is in state 3, then, there is no need to check its stability.

\subsection{Outage Probability}
In fading channels, the received signal has no constant power which depends on the channel and can be described by probability models. Thus, SINR will also become a random variable. And so the maximum capacity of the channel becomes a random variable. Outage probability is a metric for the channel that states according to the variable SINR at the received end, what is the probability that a rate is not supported due to variable SINR. In other words, outage probability at given SINR thresholds are defined as the probability of SINR value being smaller than a given threshold.
\begin{equation}
Outage Probability = Pr\left(SINR < SINR_{Threshold}\right)
\end{equation}
Where, SINR is computed as in (\ref{one}). Where P is the desired power received at receiver, $I_{i}$ is the interference power received from interferer i at the receiver and $N_{0}$ is the noise power.
\begin{equation}\label{one}
SINR =\cfrac{P}{\sum_{i=1}^{N} I_i + N_0}
\end{equation}

In our model, we denote the probability that the total interference at time instant i is larger than $\delta_{Thr}$ at the WBAN by $P_{outage}$. Then, we calculate this probability by the following formula:
\begin{equation}
P_{outage} = \left(\displaystyle\sum_{j=1}^{N} \delta_{j} > \delta_{Thr} \right)
\end{equation}
We present a simplet probabilistic approach which we prove analytically it lowers the outage probability. As mentiond above, any sensor whose received SINR at the $WBAN$ is lower than a threshold is added to the interference set of sources (IS). 

We denote by $\delta_{j}$ the received SINR at a sensor in the WBAN. If $\delta_{j} < \delta_{Thr}$, an orthogonal channel is assigned to that sensor with certain probability which equals $\frac{\delta_{i}}{\delta_{Thr}}$. Thus, at time instant i, we can calculate the average interference level using the proposed probabilistic approach as follows:
\begin{equation}\label{avgsinr}
\delta_{i} = \displaystyle\sum_{j=1}^{IS} \delta_{j} \left(1 - \frac{\delta_{j}}{\delta_{Thr}}\right)
\end{equation}

Based on the probabilistic approach, any sensor with probability $\frac{\delta_{i}}{\delta_{Thr}}$ is assigned with an orthogonal channel.

\textbf{Lemma 1:} We denote by $P_{Probilistic}$ and $P_{Original}$ the outage probability of probabilistic approach and the outage probability of the original scheme respectively. Then, $P_{Probilistic}$ < $P_{Original}$.

\textbf{Proof:} Based on outage probability definition, we have:
\begin{equation} 
P_{Probilistic} = p \left(\left(\displaystyle\sum_{i=1}^{IS} \delta_{i} \left(1 - \frac{\delta_{i}}{\delta_{Thr}} \right) \right) > \delta_{Thr} \right)
\end{equation}
\begin{equation} 
 = p \left(\left(\displaystyle\sum_{i=1}^{IS} \delta_{i} - \displaystyle\sum_{i=1}^{IS}\frac{\delta_{i}^{2}}{\delta_{Thr}}   \right) > \delta_{Thr} \right)
\end{equation}
\begin{equation} 
= p \left(\displaystyle\sum_{i=1}^{IS} \delta_{i} > \delta_{Thr} + \displaystyle\sum_{i=1}^{IS} \frac{\delta_{i}^2}{\delta_{Thr}}\right) 
\end{equation}  
\begin{equation}             
< p  \left( \displaystyle\sum_{j=1}^{IS} \delta_{i} >  \delta_{Thr} \right) = P_{Original}  
\end{equation}
Where the last line of $P_{Probilistic}$ is based on the fact that the CDF is an increasing function of its argument. Therefore, the lemma is proved.

\IncMargin{1em}
\begin{algorithm}
\footnotesize
\SetKwData{Left}{left}\SetKwData{This}{this}\SetKwData{Up}{up}
\SetKwFunction{Union}{Union}\SetKwFunction{FindCompress}{FindCompress}
\SetKwInOut{Input}{input}\SetKwInOut{Output}{output}
\Input{ISBR, C}

     $C$ $Broadcasts$ $Beacon$ $b_{k}$
     
     m = 0;
            
    \For{$i\leftarrow 1$ $\KwTo$  $sizeof(ISBR)$}
       {
     
          \If {C Acknowledges $rs_{i}$} 
            {
          
              C includes $ID_{rs_{i}}$ in nodeSlotList of $Beacon$ $b_{k + 1}$
              
              m = m + 1;
          
             }
        
       }

       C forms $nextBeacon$ $b_{k + 1}$ of p slots
       
       nodeSlotList of m slots
       
       freeSlotList of p-m slots

\caption{Coordinator Actions and Transmission Phase}
\label{cactions}
\end{algorithm}
\DecMargin{1em}
\section{Simulation Results}
We assume N=12 sources, R=4 relays and a coordinator C form a WBAN. We consider a sensor node scans different channels and eventually uses one channel at a time. We also consider that all sensor nodes excluding the coordinator in the WBAN have similar transmit power. However, in the following subsections, we prove that our proposed CFTIM scheme significantly mitigates intra-WBAN interference. Thus, the simulation results also prove that the WBAN energy savings and the throughput are enhanced. The following \textbf{table} \ref{tab:title} summarizes the simulation parameters.

\subsection{Signal to Interference and Noise Ratio (SINR)}
\begin{figure}[ht]
  \centering
        \includegraphics[width=0.4\textwidth]{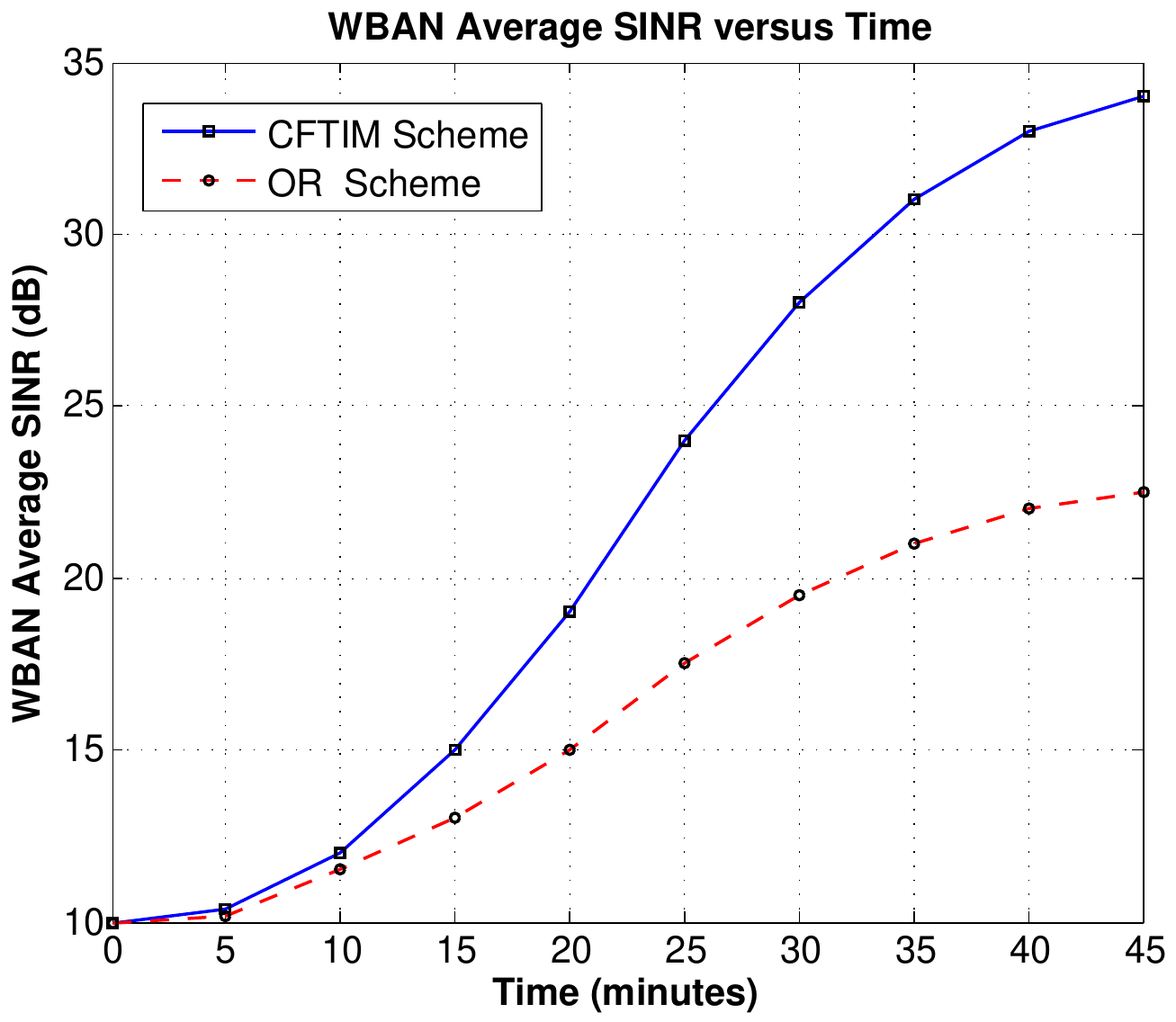}
\caption{WBAN average SINR versus time of the proposed CFTIM scheme compared to that of OR scheme}
  \label{sinrcftim}
\end{figure}

The WBAN average SINR is defined as in \textbf{Eq. \ref{avgsinr}} of the proposed CFTIM compared to that of opportunistic relaying (OR) scheme is shown in Fig. \ref{sinrcftim}.
As can be clearly seen from the figure, both curves increase with time. Also, the curve of the proposed CFTIM scheme is always higher than that of opportunistic relaying (OR) scheme \cite{key7}. Which implies, CFTIM better mitigates WBAN interference than OR scheme. Starting from average SINR of 10dB, after 25 minutes, both CFTIM and OR schemes have respectively 24dB and 17dB (after 45 minute respectively 34dB and 23dB). Therefore, the improvement in SINR from OR to CFTIM scheme is around 11dB which is quite good in WBANs. However, in the OR scheme, the interferences can happen due to the collisions at the relays or when some sources access the shared channel at the same time.  Consequently, few sources transmit and the rest will defer their transmissions (interfering sources). Therefore, these interfering nodes, with CFTIM scheme, are assigned dynamically time slots and stable channels to avoid interference with other nodes. Doing so, The average SINR is improved better in our proposed CFTIM scheme.

\subsection{WBAN Energy Consumption}
\begin{figure}[ht]
  \centering
        \includegraphics[width=0.4\textwidth]{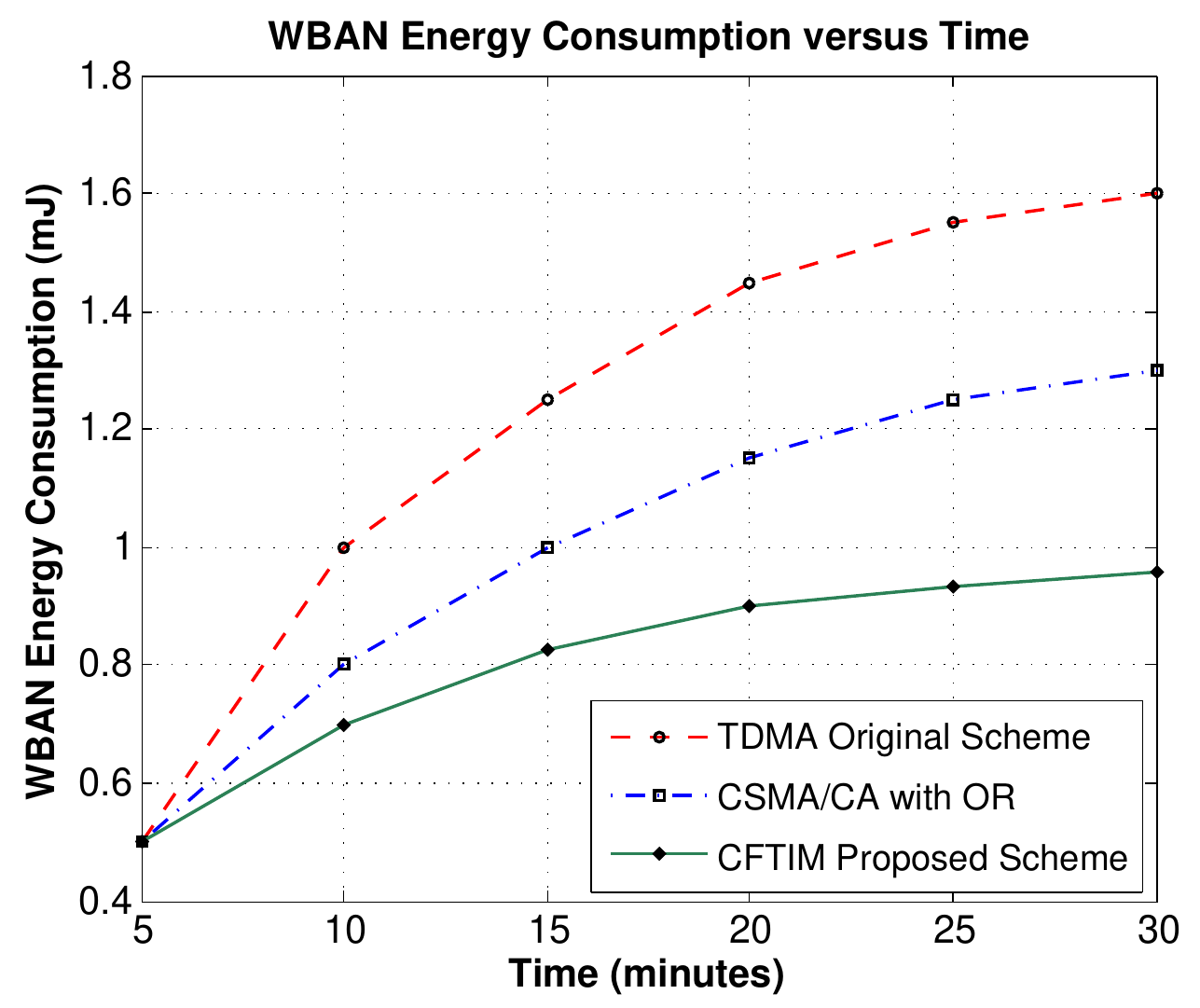}
\caption{WBAN energy consumption versus time of proposed CFTIM scheme compared to that of traditional TDMA and OR schemes}
  \label{totalenergy}
\end{figure}

In this section, the WBAN energy consumption of the proposed CFTIM scheme is thoroughly analyzed and compared with two other schemes. The first WBAN employs CSMA/CA OR communication scheme and the second employs the traditional single link TDMA scheme. The WBAN energy consumption versus time for all WBANs employing different schemes is shown in Fig. \ref{totalenergy}. As can be seen in this figure, higher WBAN energy consumption is obtained in the case of original TDMA scheme. This is because TDMA scheme uses the single-hop communication which increases the energy consumption. 
However, a lower WBAN energy consumption is obtained in the case of OR than the original TDMA scheme. The improvement in energy consumption is due to the fact that using two-hop communication reduces the energy consumption of the whole WBAN. Furthermore, it is clear to see from the figure that the proposed CFTIM scheme achieves the lowest energy consumption among others schemes. Firstly, using two-hop scheme saves the WBAN energy better than the TDMA scheme. Secondly, Avoiding interferecne using flexible TDMA at the interfering sources and the relays reduces the collisions and retransmissions and so decreases more the energy consumption.

\subsection{Throughput}
\begin{figure}[ht]
  \centering
        \includegraphics[width=0.4\textwidth]{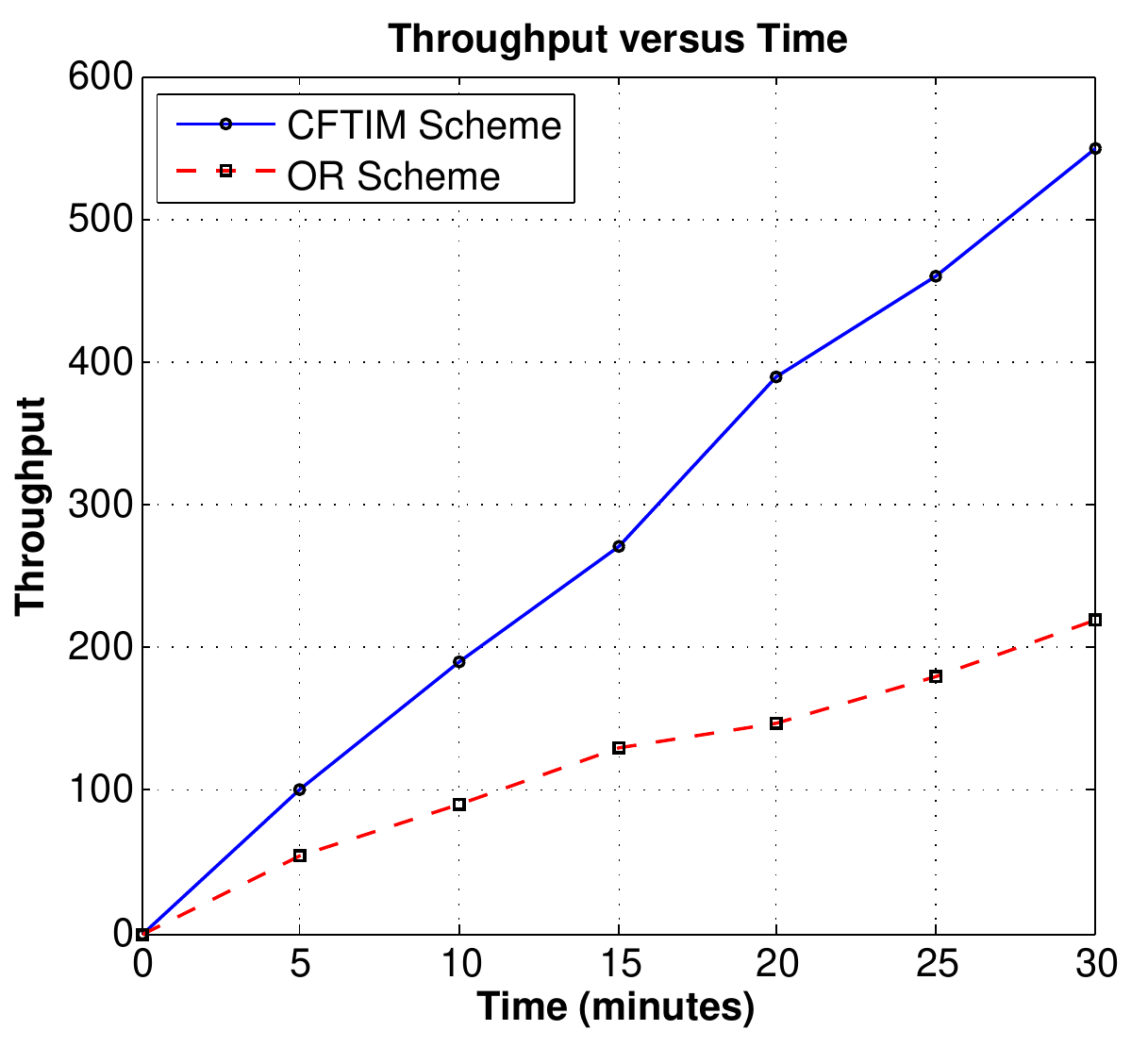}
\caption{Throughput results of the proposed CFTIM scheme versus Time in comparison to that of OR scheme}
  \label{fig:throughput}
  \end{figure}

We define the throughput as the sum of the number of successful messages delivered per a unit time at a node. The throughput versus time is shown in Fig. \ref{fig:throughput}. 

This figure compares the throughput results of our proposed CFTIM with that of OR scheme received at the coordinator. As can be clearly seen from Fig. \ref{fig:throughput} that the throughput achieved from our CFTIM scheme is always higher than that of OR scheme. This improvement can be explained due to the fact that the interference is significantly avoided in CFTIM scheme through using flexible TDMA combined with stable channel mechanism. It is evident to notice that those nodes which were supposed to experience interference in the OR scheme are avoided with CFTIM scheme which enhances the SINR and in turn improves the throughput. 

\section{Conclusion}
In this work, a distributed combined CSMA/CA with flexible TDMA scheme namely, CFTIM is proposed for interference mitigation in relay-assisted intra-WBAN. In our proposed method, non interfering sources use CSMA/CA to communication with the relays. Whilst high interfering sources together with  best relays use a flexible TDMA integrated with stable channel mechanism to communication to C. Our approach aims to minimize intra-WBAN interference. Our proposed scheme has been evaluated in comparison with other schemes. The simulation results show that the interference and the WBAN energy consumption are significantly minimized and the throughput is increased. Additionally, simple theoretical analysis of outage probability validated our proposed CFTIM approach in interference mitigation. 
  \begin{table}
\centering
\normalsize
 \begin{tabular}{ |p{3cm}|p{3cm}|  }
\hline
Simulation time & 45 minutes \\\hline
Transmission power & 0 dBm \\\hline
Noise floor & -100 dBm \\\hline
Data rate & 250 kbps  \\\hline
Packet size & 12 bytes \\\hline
Frequency & 2.4 GHz  \\\hline
Pathloss exp. ($\alpha$) &  4.22 \\\hline
Channels & 8 \\\hline
\end{tabular}
 \caption {Simulation Parameters} \label{tab:title}
\end{table}


\end{document}